\documentclass{aa}

\usepackage{upgreek}
\usepackage{txfonts}
\usepackage{graphicx}
\usepackage{xcolor}
\usepackage[colorlinks=true, linkcolor=purple, citecolor=purple, urlcolor=purple]{hyperref}
\usepackage{natbib}

\bibpunct{(}{)}{;}{a}{}{,} 

\newcommand{\bfx}{{\boldsymbol{x}}}
\newcommand{\bfv}{{\boldsymbol{\upsilon}}}
\newcommand{\txa}{{\text{a}}}

\newcommand{\txm}{{\text{m}}}
\newcommand{\txd}{{\text{d}}}

\newcommand{\calE}{{\cal{E}}}

\begin{document}

\title{The differential energy distribution \\and the total integrated binding energy of dynamical models}
\titlerunning{Total integrated binding energy}

\author{Maarten Baes \and Herwig Dejonghe}
\authorrunning{Baes \& Dejonghe}

\date{Received 3 June 2021 / Accepted 23 July 2021}

\institute{
Sterrenkundig Observatorium, Universiteit Gent, Krijgslaan 281 S9, 9000 Gent, Belgium\\ e-mail: {\tt{maarten.baes@ugent.be}}
\label{UGent}
}

\abstract{%
We revisit the differential energy distribution of steady-state dynamical models. It has been shown that the differential energy distribution of steady-state spherical models does not vary strongly with the anisotropy profile, and that it is hence mainly determined by the density distribution of the model. We explore this similarity in more detail. Through a worked example and a simple proof, we show that the mean binding energy per unit mass $\langle\calE\rangle$, or equivalently the total integrated binding energy $B_{\text{tot}} = M\langle\calE\rangle$, is independent of the orbital structure, not only for spherical models but for any steady-state dynamical model. Only the higher-order moments of the differential energy distribution depend on the details of the orbital structure. We show that the standard deviation of the differential energy distribution of spherical dynamical models varies systematically with the anisotropy profile: radially anisotropic models tend to prefer more average binding energies, whereas models with a more tangential orbital distribution slightly favour more extreme binding energies. Finally, we find that the total integrated binding energy supplements the well-known trio consisting of total kinetic energy, total potential energy, and total energy on an equal footing. Knowledge of any one out of these four energies suffices to calculate the other three.}


\maketitle

\section{Introduction}

In stellar dynamics, the most fundamental quantity of a self-gravitating system is the phase space distribution function $f(t,\bfx,\bfv)$. It describes the density of stars in the six-dimensional phase space and contains all the dynamical information. As the mass density $\rho(t,\bfx)$ is obtained by integrating $f(t,\bfx,\bfv)$ over velocity space, there are an infinite number of dynamical models with different orbital structures that correspond to a given density. In stationary spherical symmetry, various techniques to generate dynamical models with a different anisotropy profile for a given density profile have been presented, for example by \citet{1979PAZh....5...77O}, \citet{1984ApJ...286...27R}, \citet{1984A&A...133..225D, 1986PhR...133..217D, 1989ApJ...343..113D}, \citet{1985AJ.....90.1027M}, \citet{1991MNRAS.253..414C}, \citet{1991MNRAS.250..812G}, and \citet{1995MNRAS.275.1017C}. Well-known examples of spherical models for which different analytical distribution functions have been derived include the Plummer, Hernquist, and Jaffe models \citep{1983MNRAS.202..995J, 1985AJ.....90.1027M, 1985MNRAS.214P..25M, 1987MNRAS.224...13D, 1990ApJ...356..359H, 1991MNRAS.253..414C, 2007A&A...471..419B}. 

It is customary to ignore possible contributions from matter that is not bound to the stellar system under consideration, hence it is convenient to adopt the symbol ${\cal E}=-E\ge0$ for the binding energy per unit mass. A characteristic of dynamical models that has drawn some attention is the differential energy distribution $N(\calE)$, that is, the distribution of mass as a function of $\calE$. It is a natural diagnostic for dynamical models that is easily calculated from $N$-body simulations \citep[e.g.][]{1982MNRAS.201..939V, 2001ApJ...554.1268H, 2013MNRAS.431.3177D, 2020MNRAS.491.4591E}. It has been argued that the differential energy distribution is a fundamental partitioning of a stellar system, as equilibrium stellar dynamical systems are collisionless systems in which all particles retain their energies \citep{1982MNRAS.200..951B, 2007LNP...729..297E, 2010ApJ...722..851H}.

Given that for a fixed mass density dynamical models with a widely varying orbital structure can be generated, one would expect that these different models would also have strongly different differential energy distributions. The opposite turns out to be the case, however, in different case studies of spherical models. \citet{1987gady.book.....B} considered two different models with the same Jaffe density profile, an isotropic model, and a model consisting of only radial orbits, and they showed that they have a very similar differential energy distribution. \citet{1987MNRAS.224...13D} presented a family of Plummer models with an isotropic anisotropy profile in the central regions, but strong anisotropy in the outer regions. He concluded that the differential energy distribution of the models in this family did not differ drastically, even when comparing models that range from completely radial to strongly tangentially anisotropic in the outer regions. A similar conclusion was drawn by \citet{1995MNRAS.275.1017C} based on a similar family of anisotropic Plummer models. Finally, \citet{2008gady.book.....B} presented the differential energy distribution for three spherical models with a Hernquist density profile, but with different anisotropy profiles: a radially anisotropic, an isotropic, and a tangentially anisotropic model. While the distribution functions of these models are very different, their differential energy distributions are, again, very similar. 

All of these studies came to a similar qualitative conclusion, namely that the differential energy distribution generally does not seem to vary strongly with the anisotropy profile and hence mainly depends on the mass distribution. However, no satisfactory explanation has been provided for this systematic trend. The goal of the present paper is to quantify the similarity of the differential energy distribution of steady-state dynamical models in more detail. In Sec.~2 we analyse the differential energy distribution for a family of Hernquist models that covers an extreme range in the anisotropy profile (ranging from completely radial to completely circular orbits). We show that all these models have the same mean binding energy per unit mass, which contributes to the similarity of their differential energy distributions. In Sec.~3 we generalise this result for all steady-state dynamical models corresponding to a given mass density, and we derive a general relation between the total integrated binding energy and other measures for the total energy content. In Sec.~4 we discuss our results, and Sec.~5 contains a summary.

\section{The Hernquist model}

We start our analysis with a simple but illustrative example: the Hernquist model. This model, introduced by 
\citet{1990ApJ...356..359H}, has a very simple density profile 
\begin{equation}
\rho(r) = \frac{M}{2\pi}\,\frac{b}{r\,(b+r)^3},
\end{equation}
where $M$ represents the total mass and $b$ a scale radius. The (positive) gravitational potential is 
\begin{equation}
\Psi(r) = \frac{GM}{b+r}.
\end{equation}
The total potential energy of the Hernquist model is \citep{1990ApJ...356..359H, 2019A&A...630A.113B},
\begin{equation}
\label{Wtot}
W_{\text{tot}} = -\frac{GM^2}{6b}.
\end{equation}

\subsection{Constant anisotropy models}

\citet{2002A&A...393..485B} discussed different dynamical models that all generate the Hernquist model. They considered, amongst others, a one-parameter family of models with constant anisotropy $\beta$. In Appendix~{\ref{AppA}} we derive explicit expressions for the differential energy distribution of this family of models. We note that this family covers a very wide range of orbital structure, ranging from a model with all stars on purely radial orbits $(\beta=1$), over an isotropic model ($\beta=0$), to a model with all stars on circular orbits ($\beta=-\infty)$. We note that in accordance with the density slope--anisotropy relation \citep{2006ApJ...642..752A, 2010MNRAS.408.1070C, 2011ApJ...726...80V}, only models with $\beta\leqslant\tfrac12$ have positive distribution functions, which means that models with $\beta>\tfrac12$ are nonphysical. We can, however, still formally derive the differential energy distribution for models with $\beta>\tfrac12$, but these are all negative at large binding energies.

\begin{figure*}
\includegraphics[width=0.97\textwidth]{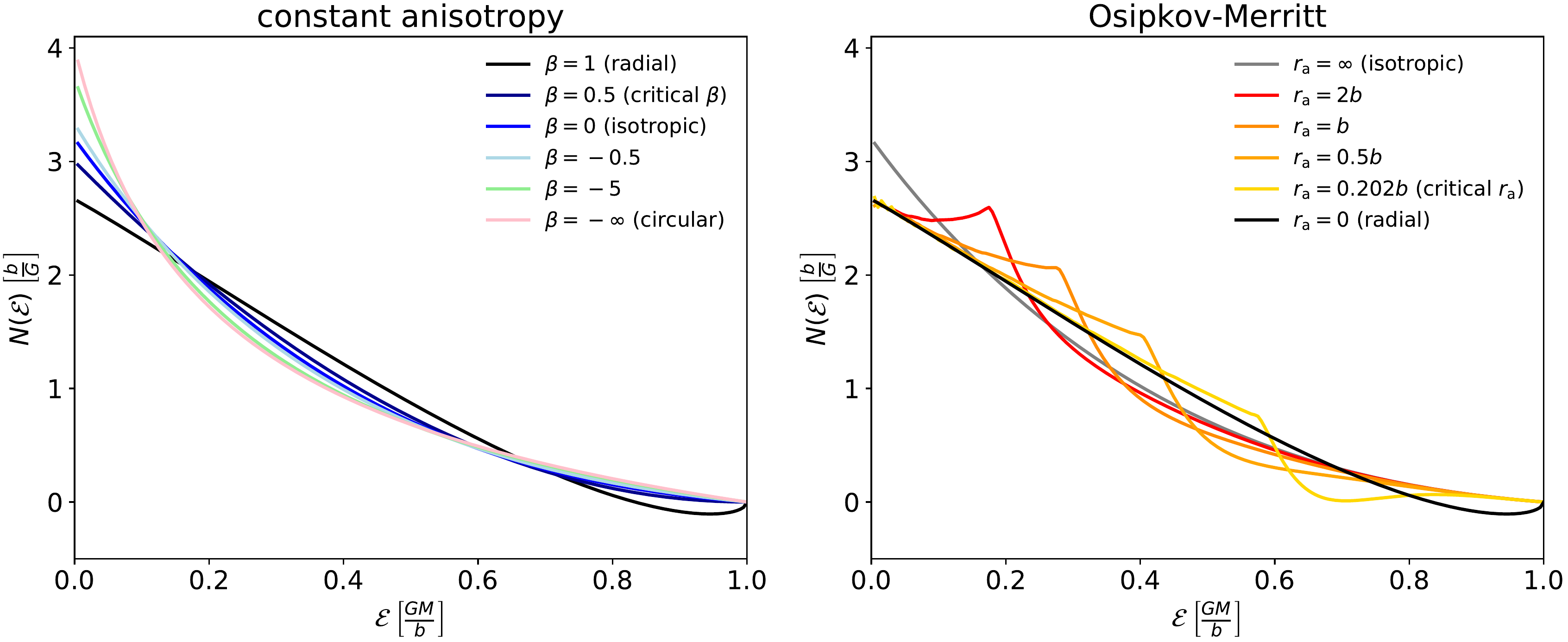}%
\caption{Differential energy distributions for Hernquist models with different anisotropy profiles. Left panel: models with a constant anisotropy for different values of the anisotropy parameter $\beta$. Right panel: models with an Osipkov-Merritt anisotropy profile for different values of the anisotropy radius $r_\txa$. We note that the discontinuity in the Osipkov-Merritt distribution function of these models leaves its mark.}
\label{H-DED.fig}
\end{figure*}

In the left panel of Fig.~{\ref{H-DED.fig}} we plot the differential energy distributions for different members of this family of constant anisotropy Hernquist models. The models corresponding to $\beta=\tfrac12$, $\beta=0$, and $\beta=-\tfrac12$ are the same models as shown in Fig.~{4.5} of  \citet{2008gady.book.....B}, Fig.~{\ref{H-DED.fig}} covers a wider range in the anisotropy profile. 

The first obvious conclusion is that the differential energy distributions are very similar, in spite of the very different anisotropy of the models. Inspecting the different curves more closely, we note that the shape of the differential energy distribution changes in a systematic way as the anisotropy of the model changes from purely radial to purely circular orbits. $N(\calE)$ is a decreasing function of $\calE$ at both low and high binding energies, and an increasing function of $\beta$ at intermediate binding energies. In other words, tangentially anisotropic models seem to have a slight preference for stars on orbits with more extreme binding energies, whereas radial models slightly prefer average binding energies. On average, the different differential energy distributions seem to have the same mean value. 

This qualitative impression can be quantified  by looking at the moments of the differential energy distribution. As the differential energy distribution is just the distribution of mass as a function of binding energy, we obviously must have 
\begin{equation}
\int_0^{\Psi_0} N(\calE)\,\txd\calE = M,
\label{H-norm}
\end{equation}
with $\Psi_0$ the depth of the potential well and $M$ the total mass of the system. All of the differential energy distributions presented in Appendix~{\ref{AppA}} and shown in the left panel of Fig.~{\ref{H-DED.fig}} satisfy this normalisation. Even the non-physical differential energy distribution (\ref{dedrad}) of the hypothetical purely radial model satisfies this normalisation.

Next, we look at the mean binding energy per unit mass,
\begin{equation}
\langle\calE\rangle = \frac{1}{M} \int_0^{\Psi_0} N(\calE)\,\calE\,\txd\calE.
\end{equation}
For all the constant anisotropy Hernquist models, $\langle\calE\rangle$ turns out to have the same value, irrespective of the value of $\beta$,
\begin{equation}
\langle\calE\rangle = \frac{GM}{4b}.
\label{H-meanE}
\end{equation}
This is an interesting finding: in spite of the fact that the anisotropy can be extremely different, ranging from purely radial to purely circular orbits, the mean binding energy is exactly the same. We can express this result in an equivalent way by introducing the total integrated binding energy $B_{\text{tot}}$ as
\begin{equation}
B_{\text{tot}} \equiv M\langle\calE\rangle = \int_0^{\Psi_0} N(\calE)\,\calE\,\txd\calE.
\end{equation}
Taking into account expression (\ref{Wtot}) we find for the family of constant anisotropy Hernquist models the relation
\begin{equation}
B_{\text{tot}} = -\tfrac32\,W_{\text{tot}}. 
\label{H-Btot}
\end{equation}
The total integrated binding energy is hence independent of the value of $\beta$. 

Since the normalisation and the mean value of all differential energy distributions are identical, deviations between them must be found in the higher-order moments only. The standard deviation of the differential energy distribution is calculated as
\begin{equation}
\sigma_\calE = \left[\frac{1}{M} \int_0^{\Psi_0} N(\calE)\, \Bigl(\calE - \langle\calE\rangle \Bigr)^2\,\txd\calE \right]^{1/2}.
\end{equation}
For the family of Hernquist models with a constant anisotropy, the standard deviation decreases slightly but systematically with increasing $\beta$, ranging from 0.0458 for the model with purely circular orbits, over 0.0375 for the boundary model with $\beta=\tfrac12$, to 0.0299 for the nonphysical model with only radial orbits (all in units $GM/b$). This confirms the systematic shift from a slight excess for extreme binding energies for tangential models to a preference for average binding energies for radially anisotropic models. 

\subsection{Osipkov-Merritt models}

In order to exclude the possibility that the property (\ref{H-meanE}) is a special property of models with a constant anisotropy, we also calculate the differential energy distribution corresponding to Hernquist models with an Osipkov-Merritt anisotropy profile \citep{1979PAZh....5...77O,1985AJ.....90.1027M}. These models are characterised by an anisotropy parameter $\beta(r)$ that increases smoothly from 0 at small radii to 1 at large radii, that is the models are isotropic in the centre and completely radially anisotropic in the outskirts. The transition from isotropic to radial is characterised by an anisotropy radius $r_\txa$. The distribution function for the Osipkov-Merritt Hernquist model can be calculated analytically \citep{1990ApJ...356..359H, 2002A&A...393..485B}, but the differential energy distribution needs to be determined numerically. 

The result is shown in the right panel of Fig.~{\ref{H-DED.fig}} for different values of the anisotropy parameter. This plot shows a systematic trend when the anisotropy radius increases from $r_\txa = 0$, corresponding to the hypothetical model with only radial orbits, to $r_\txa=\infty$, corresponding to the isotropic model. The differential energy distributions all asymptotically behave as the pure radial model at low binding energies, and they asymptotically approach the isotropic model at high binding energies. The larger the anisotropy radius, the more the differential energy distribution approximates the one corresponding to the isotropic model. 

We have also numerically calculated the moments of the differential energy distribution for these Osipkov-Merritt models. First of all, all the differential energy distributions obviously satisfy the normalisation (\ref{H-norm}), which is good validation test for the accuracy of our calculations. Secondly, it turns out that they also all satisfy the relation (\ref{H-meanE}), or equivalently relation (\ref{H-Btot}), that is, the mean binding energy per unit mas is independent of the anisotropy radius. The fact that we find the same relation as for the constant anisotropy models strongly suggests that the mean binding energy per unit mass, or equivalently, the total integrated binding energy, of all Hernquist models is exactly the same, independent of the anisotropy profile.
 
Finally, we can also quantify the difference between the different differential energy distributions through the standard deviation. We find that, for the Osipkov-Merritt Hernquist models, $\sigma_\calE$ increases systematically from 0.0299 to 0.0403, in units of $GM/b$, when the anisotropy radius increases from $r_\txa=0$ (model with only radial orbits) to $r_\txa=\infty$ (isotropic model). This is a similar trend as for the constant anisotropy models: more radially anisotropic models tend to have smaller values of $\sigma_\calE$, or in other words, to prefer more average binding energies.

\section{The total integrated binding energy of equilibrium dynamical models}

The analysis in the previous section suggests that the mean binding energy per unit mass $\langle\calE\rangle$, or equivalently the total integrated binding energy $B_{\text{tot}}$, of spherical dynamical models with a fixed density profile is independent of the details of the anisotropy profile. In this section we show that this hypothesis is true, not only for spherical models but for any steady-state dynamical model. This turns out to be a simple exercise.

To prove this conjecture, we start with the general expression for the differential energy distribution provided by \citet{2008gady.book.....B},
\begin{equation}
N(\calE) = \int \txd\bfx \int \txd\bfv\, f(\bfx,\bfv)\,
\delta\left(\Psi(\bfx)-\tfrac12|\bfv|^2-\calE\right).
\label{NEgen}
\end{equation}
The total integrated binding energy is thus given by 
\begin{equation}
B_{\text{tot}}
= 
\int \calE\,\txd\calE
\int \txd\bfx \int f(\bfx,\bfv)\,
\delta\left(\Psi(\bfx)-\tfrac12|\bfv|^2-\calE\right) \txd\bfv.
\end{equation}
Permuting the order of integration yields
\begin{equation}
B_{\text{tot}}
= 
\int \txd\bfx \int f(\bfx,\bfv)\,\txd\bfv 
\int \calE\,
\delta\left(\Psi(\bfx)-\tfrac12|\bfv|^2-\calE\right)\,\txd\calE
\label{Btotint}
\end{equation}
or
\begin{multline}
B_{\text{tot}}
=  
\int \Psi(\bfx)\,\txd\bfx \int f(\bfx,\bfv)\,\txd\bfv
\\
-\frac12 \int \txd\bfx \int f(\bfx,\bfv)\,|\bfv|^2\,\txd\bfv. 
\qquad
\end{multline}
The inner integral in the first term on the right-hand side of this equation equals the mass density, whereas the integral in the second term is the total kinetic energy:
\begin{equation}
B_{\text{tot}}
= 
\int \rho(\bfx)\,\Psi(\bfx)\,\txd\bfx 
-\frac12 \int \rho(\bfx)\,\langle \upsilon^2\rangle(\bfx)\,\txd\bfx.
\end{equation}
The first term is the total potential energy of an equilibrium dynamical system \citep{2008gady.book.....B}. With the notations
\begin{gather}
W_{\text{tot}} = -\frac12 \int \rho(\bfx)\,\Psi(\bfx)\,\txd\bfx,
\label{Wtot-def}
\\
T_{\text{tot}} = \frac12 \int \rho(\bfx)\,\langle \upsilon^2\rangle(\bfx)\,\txd\bfx,
\label{Ttot-def}
\end{gather}
we obtain for the total integrated (binding) energy
\begin{equation}
B_{\text{tot}} = -2\,W_{\text{tot}} - T_{\text{tot}}.
\label{BWT}
\end{equation}
Invoking the virial theorem, which states that $W_{\text{tot}} + 2\,T_{\text{tot}} = 0$ or $E_{\text{tot}}= W_{\text{tot}} + T_{\text{tot}}=-T_{\text{tot}}<0$, with $E_{\text{tot}}$ the total energy, we ultimately obtain the expression
\begin{equation}
B_{\text{tot}} = 3\,T_{\text{tot}} = -\tfrac32\, W_{\text{tot}}=-3\,E_{\text{tot}}.
\label{meanE}
\end{equation}
This proves our hypothesis that the total integrated binding energy of a steady-state dynamical model is independent of the details of the orbital structure. We also note that the casual argument $B_{\text{tot}}=-W_{\text{tot}} - T_{\text{tot}}=-E_{\text{tot}}$ is erroneous.

\section{Discussion}

\subsection{The similarity of the differential energy distributions}

The original starting point of this research was the observation in previous work that the  differential energy distribution of steady-state spherical dynamical models generally does not seem to vary strongly with the anisotropy profile and hence mainly depends on the mass distribution. While \citet{1987gady.book.....B, 2008gady.book.....B}, \citet{1987MNRAS.224...13D}, and \citet{1995MNRAS.275.1017C} all concluded that models with fixed density profile all have a similar differential energy distribution, the level of similarity has not been quantified. 

The first main conclusion of this work is a quantification of this statement: we demonstrate that all dynamical models corresponding a fixed density distribution have the same mean binding energy per unit mass. Since the normalisation and the mean value of the differential energy distribution of any dynamical model consistent with a given density distribution is identical, deviations between them must be found in the higher-order moments only. This explains, at least to some degree, why the distributions are so similar, in spite of the fact that the orbital structure can vary from completely circular to completely radial orbits. 

To the best of our knowledge, this simple fact has never been demonstrated. \citet{1987MNRAS.224...13D} does discuss $\mu_\calE(r)$, the mean binding energy per unit mass as a function of radius, in his analysis of a set of anisotropic Plummer models.  Fig.~2 from that paper shows how $\mu_\calE(r)$ changes as a function of the anisotropy of the models (parameterised by a parameter $q$). At large radii, models with a tangential anisotropy have a slightly smaller mean binding energy per unit mass than models with a radial anisotropy, and vice versa at small radii. This anisotropy dependence is also clear from the explicit formula for $\mu_\calE(r)$, as written in Eq.~(28) of \citet{1987MNRAS.224...13D}. Integrating this quantity over the entire physical space results in what we denote as $\langle\calE\rangle$. This integral results in $\langle\calE\rangle = 9\pi/64$, independent of $q$ as required.

\subsection{Systematic differences in the differential energy distributions}

The second general conclusion from our work relates to these second-order differences between the differential energy distributions corresponding to spherical models with different anisotropy profiles. Both the constant anisotropy and the Osipkov-Merritt Hernquist models presented in Sec.\ 2 showed a systematic behaviour between the anisotropy profile and the shape of the differential energy, as parametrised by the standard deviation. The bottomline is that more radially anisotropic models tend to prefer more average binding energies, whereas models with a more tangential orbital distribution slightly favour more extreme binding energies. We note that these differences always need to be symmetric, because the mean binding energy is independent of the anisotropy profile. 

Inspecting previous work that discussed the differential energy distribution as a function of the anisotropy profile, the same trend is seen. It is most clearly present in the two sets of Plummer models discussed by \citet{1987MNRAS.224...13D} and \citet{1995MNRAS.275.1017C}. As shown in their Fig.~3 and Fig.~3a, respectively, the models with a radial anisotropy have a clearly peaked differential energy distribution with a maximum around $\langle\calE\rangle = 9\pi/64$, whereas the differential energy distribution of tangential models is much broader and less strongly peaked. For the most strongly tangential models shown by \citet{1995MNRAS.275.1017C}, the distribution is even decidedly boxy.

The reason for this systematic shift needs to be sought in the different contribution of potential and kinetic energy to the binding energy. In a dynamical model mainly consisting of radial orbits, many particles, especially at large radii, spend a considerable fraction of their time near the apocentre. Their kinetic energy is modest, so the binding energy corresponding to these orbits is roughly equal to the (positive) potential energy, with almost no kinetic energy subtracted. On the other hand, since all particles on strongly radial orbits also cross the inner regions, there is little additional room for highly bound particles on tangential orbits in the central regions. The final result is that radially anisotropic models contain few particles with either very small or very large binding energies, and thus display a preference for average binding energies. Strongly tangential models on the other hand show the opposite behaviour: they contain both tightly bound particles that remain in the very inner regions, whereas the particles in the outer regions still have a significant kinetic energy, almost at the same magnitude of the potential energy. The result is that tangentially anisotropic models prefer more extreme binding energies. Interestingly, this excess at both small and large binding energies is nicely balanced such that the average binding energy is independent of the anisotropy profile. 

\subsection{The total integrated binding energy}

The last result from our study concerns the actual value of the $\langle\calE\rangle$, or equivalently of the total integrated binding energy $B_{\text{tot}} = M\langle\calE\rangle$. It turns out to be quite straightforward to demonstrate that, for any steady-state dynamical model, $B_{\text{tot}}$ can be written as expression~(\ref{BWT}). We can write $B_{\text{tot}}$ in this form due to the simple fact that the binding energy per unit mass is the sum of the binding potential and the kinetic energy per unit mass, that is, it is a linear combination of these two contributions. Any higher-order moment of the differential energy distributions contains mixed combinations of these contributions. Interestingly, we now have three different linear combinations of the total potential and total kinetic energy of a steady-state dynamical model:
\begin{gather}
2\,W_{\text{tot}} + T_{\text{tot}} = -B_{\text{tot}},
\\
W_{\text{tot}} + T_{\text{tot}} = E_{\text{tot}},
\label{Etot}
\\
W_{\text{tot}} + 2\,T_{\text{tot}} = 0,
\end{gather}
which leads to 
\begin{equation}
B_{\text{tot}} = 3\,T_{\text{tot}} = -\tfrac32\, W_{\text{tot}} = -3\,E_{\text{tot}}.
\label{energyrelation}
\end{equation}
The total integrated binding energy thus supplements the well-known trio consisting of total kinetic energy, total potential energy, and total energy on equal footing. The four energies $(B_{\text{tot}}, T_{\text{tot}}, W_{\text{tot}}, E_{\text{tot}})$ form a quadruple, in which knowledge of any one out of these four suffices to calculate the other three. This result is the more remarkable in view of the fact that knowledge of $N(\calE)$, say from an $N$-body calculation, would not seem to suffice in order to calculate the total kinetic and/or potential energies separately. Clearly it does. 

These relations can be extended to the case of alternative theories of gravity where the force differs from the Newtonian $r^{-1}$ potential. The main result, expression~(\ref{BWT}), does not depend on the nature of the potential. In the case of an interparticle power-law potential of the form $r^{-p}$, as studied by many previous authors \citep[e.g.][]{2001PhRvE..64e6103I, 2002PhRvE..66e1112I, 2013MNRAS.431.3177D, 2015JPlPh..81e4904D, 2017MNRAS.468.2222D, 2017PhRvE..96c2102M}, the scalar virial theorem can be written as \citep[][Problem 7.1]{2008gady.book.....B}
\begin{equation}
p\,W_{\text{tot}} + 2\,K_{\text{tot}} = 0.
\end{equation}
Combining this with the expressions (\ref{BWT}) and (\ref{Etot}) leads to a generalisation of Eq.~(\ref{energyrelation}),
\begin{equation}
B_{\text{tot}} = \left(\frac{4-p}{p}\right)T_{\text{tot}} = -\left(\frac{4-p}{2}\right) W_{\text{tot}} = -\left(\frac{4-p}{2-p}\right)E_{\text{tot}}.
\end{equation}

\section{Summary}

The starting point for this paper is the qualitative observation that the differential energy distribution of steady-state spherical models does not vary strongly with the anisotropy profile. Our investigation, based on a detailed study of a family of Hernquist models and a rigorous general proof, has yielded the following conclusions. 

Firstly we demonstrate that all dynamical models corresponding a fixed density distribution have the same mean binding energy per unit mass. Differences between differential energy distributions are only visible in the higher-order moments. This helps to explain why the distributions are so similar, even for orbital structures ranging from completely circular to completely radial orbits. 

Secondly we show that the shape of the differential energy distribution changes systematically with the anisotropy profile for spherical dynamical models with a given density distribution, radially anisotropic models tend to prefer more average binding energies, whereas models with a more tangential orbital distribution slightly favour more extreme binding energies. These differences are always symmetric because the mean binding energy is independent of the anisotropy profile. This systematic effect can be understood as the result of the different contribution of potential and kinetic energy to the binding energy. 

Finally, we demonstrate that the total integrated binding energy supplements the well-known triplet consisting of total kinetic energy, total potential energy, and total energy on an equal footing. The four energies form a quadruple, in which knowledge of any one out of these four suffices to calculate the other three.

\bibliography{DED}

\begin{thebibliography}{35}
\expandafter\ifx\csname natexlab\endcsname\relax\def\natexlab#1{#1}\fi

\bibitem[{{An} \& {Evans}(2006)}]{2006ApJ...642..752A}
{An}, J.~H. \& {Evans}, N.~W. 2006, \apj, 642, 752

\bibitem[{{Baes} \& {Ciotti}(2019)}]{2019A&A...630A.113B}
{Baes}, M. \& {Ciotti}, L. 2019, \aap, 630, A113

\bibitem[{{Baes} \& {Dejonghe}(2002)}]{2002A&A...393..485B}
{Baes}, M. \& {Dejonghe}, H. 2002, \aap, 393, 485

\bibitem[{{Baes} \& {van Hese}(2007)}]{2007A&A...471..419B}
{Baes}, M. \& {van Hese}, E. 2007, \aap, 471, 419

\bibitem[{{Binney}(1982)}]{1982MNRAS.200..951B}
{Binney}, J. 1982, \mnras, 200, 951

\bibitem[{{Binney} \& {Tremaine}(1987)}]{1987gady.book.....B}
{Binney}, J. \& {Tremaine}, S. 1987, {Galactic Dynamics} (Princeton University
  Press, Princeton)

\bibitem[{{Binney} \& {Tremaine}(2008)}]{2008gady.book.....B}
{Binney}, J. \& {Tremaine}, S. 2008, {Galactic Dynamics: Second Edition}
  (Princeton University Press, Princeton)

\bibitem[{{Ciotti} \& {Morganti}(2010)}]{2010MNRAS.408.1070C}
{Ciotti}, L. \& {Morganti}, L. 2010, \mnras, 408, 1070

\bibitem[{{Cuddeford}(1991)}]{1991MNRAS.253..414C}
{Cuddeford}, P. 1991, \mnras, 253, 414

\bibitem[{{Cuddeford} \& {Louis}(1995)}]{1995MNRAS.275.1017C}
{Cuddeford}, P. \& {Louis}, P. 1995, \mnras, 275, 1017

\bibitem[{{Dejonghe}(1984)}]{1984A&A...133..225D}
{Dejonghe}, H. 1984, \aap, 133, 225

\bibitem[{{Dejonghe}(1986)}]{1986PhR...133..217D}
{Dejonghe}, H. 1986, \physrep, 133, 217

\bibitem[{{Dejonghe}(1987)}]{1987MNRAS.224...13D}
{Dejonghe}, H. 1987, \mnras, 224, 13

\bibitem[{{Dejonghe}(1989)}]{1989ApJ...343..113D}
{Dejonghe}, H. 1989, \apj, 343, 113

\bibitem[{{Di Cintio} {et~al.}(2013){Di Cintio}, {Ciotti}, \&
  {Nipoti}}]{2013MNRAS.431.3177D}
{Di Cintio}, P., {Ciotti}, L., \& {Nipoti}, C. 2013, \mnras, 431, 3177

\bibitem[{{Di Cintio} {et~al.}(2015){Di Cintio}, {Ciotti}, \&
  {Nipoti}}]{2015JPlPh..81e4904D}
{Di Cintio}, P., {Ciotti}, L., \& {Nipoti}, C. 2015, Journal of Plasma Physics,
  81, 495810504

\bibitem[{{Di Cintio} {et~al.}(2017){Di Cintio}, {Ciotti}, \&
  {Nipoti}}]{2017MNRAS.468.2222D}
{Di Cintio}, P., {Ciotti}, L., \& {Nipoti}, C. 2017, \mnras, 468, 2222

\bibitem[{{Eddington}(1916)}]{1916MNRAS..76..572E}
{Eddington}, A.~S. 1916, \mnras, 76, 572

\bibitem[{{Efthymiopoulos} {et~al.}(2007){Efthymiopoulos}, {Voglis}, \&
  {Kalapotharakos}}]{2007LNP...729..297E}
{Efthymiopoulos}, C., {Voglis}, N., \& {Kalapotharakos}, C. 2007, Lecture Notes
  in Physics, 729, 297

\bibitem[{{Errani} \& {Pe{\~n}arrubia}(2020)}]{2020MNRAS.491.4591E}
{Errani}, R. \& {Pe{\~n}arrubia}, J. 2020, \mnras, 491, 4591

\bibitem[{{Evans} \& {An}(2006)}]{2006PhRvD..73b3524E}
{Evans}, N.~W. \& {An}, J.~H. 2006, \prd, 73, 023524

\bibitem[{{Gerhard}(1991)}]{1991MNRAS.250..812G}
{Gerhard}, O.~E. 1991, \mnras, 250, 812

\bibitem[{{Hanyu} \& {Habe}(2001)}]{2001ApJ...554.1268H}
{Hanyu}, C. \& {Habe}, A. 2001, \apj, 554, 1268

\bibitem[{{Hernquist}(1990)}]{1990ApJ...356..359H}
{Hernquist}, L. 1990, \apj, 356, 359

\bibitem[{{Hjorth} \& {Williams}(2010)}]{2010ApJ...722..851H}
{Hjorth}, J. \& {Williams}, L. L.~R. 2010, \apj, 722, 851

\bibitem[{{Iguchi}(2002)}]{2002PhRvE..66e1112I}
{Iguchi}, O. 2002, \pre, 66, 051112

\bibitem[{{Ispolatov} \& {Cohen}(2001)}]{2001PhRvE..64e6103I}
{Ispolatov}, I. \& {Cohen}, E.~G.~D. 2001, \pre, 64, 056103

\bibitem[{{Jaffe}(1983)}]{1983MNRAS.202..995J}
{Jaffe}, W. 1983, \mnras, 202, 995

\bibitem[{{Marcos} {et~al.}(2017){Marcos}, {Gabrielli}, \&
  {Joyce}}]{2017PhRvE..96c2102M}
{Marcos}, B., {Gabrielli}, A., \& {Joyce}, M. 2017, \pre, 96, 032102

\bibitem[{{Merritt}(1985{\natexlab{a}})}]{1985MNRAS.214P..25M}
{Merritt}, D. 1985{\natexlab{a}}, \mnras, 214, 25P

\bibitem[{{Merritt}(1985{\natexlab{b}})}]{1985AJ.....90.1027M}
{Merritt}, D. 1985{\natexlab{b}}, \aj, 90, 1027

\bibitem[{{Osipkov}(1979)}]{1979PAZh....5...77O}
{Osipkov}, L.~P. 1979, Pisma v Astronomicheskii Zhurnal, 5, 77

\bibitem[{{Richstone} \& {Tremaine}(1984)}]{1984ApJ...286...27R}
{Richstone}, D.~O. \& {Tremaine}, S. 1984, \apj, 286, 27

\bibitem[{{van Albada}(1982)}]{1982MNRAS.201..939V}
{van Albada}, T.~S. 1982, \mnras, 201, 939

\bibitem[{{Van Hese} {et~al.}(2011){Van Hese}, {Baes}, \&
  {Dejonghe}}]{2011ApJ...726...80V}
{Van Hese}, E., {Baes}, M., \& {Dejonghe}, H. 2011, \apj, 726, 80

\end{thebibliography}

\appendix
\section{Differential energy distributions for the constant anisotropy Hernquist models}
\label{AppA}

For a spherical dynamical model characterised by a general distribution function $f(\calE,L)$, expression the differential energy distribution can be written as 
\begin{equation}
N(\calE) 
= 
16\pi^2
\int_0^{r_\txm(\calE)} \txd r
\int_0^{\sqrt{2r^2(\Psi-\calE)}} 
\frac{f(\calE,L)\,L\,\txd L}{\sqrt{2(\Psi-\calE)-L^2/r^2}},
\label{NE}
\end{equation}
with $r_{\text{m}}(\calE)$ the maximum radius that can be reached by a star with binding energy $\calE$, defined through $\Psi(r_\txm(\calE))=\calE$. In this Section we derive explicit expressions for the differential energy distribution for Hernquist models with a constant anisotropy, including the limiting cases of models consisting of purely radial or purely circular orbits.

\subsection{Isotropic model}

For isotropic models, the distribution function is function of $\calE$ only and can be calculated from the density by means of Eddington's formula \citep{1916MNRAS..76..572E}. If the distribution function only depends on binding energy, $f(\calE)$ can be taken outside the integrals in Eq.~(\ref{NE}) and the inner integral can be evaluated analytically. The result is
\begin{equation}
N(\calE) = 16\!\sqrt{2}\,\pi^2\,f(\calE) \int_0^{r_\txm(\calE)} \sqrt{\Psi-\calE}\,r^2\,\txd r.
\label{Hisodf}
\end{equation}
For the isotropic Hernquist model, the distribution function can be written in terms of elementary functions, and also the remaining integral in expression~(\ref{Hisodf}) can be evaluated analytically. The final result is \citep{1990ApJ...356..359H},
\begin{multline}
N(\calE) = \frac{b}{12\pi G} \,\frac{1}{\varepsilon^4\,(1-\varepsilon^2)^2} 
\\
\qquad\qquad\times
\left[\frac{3\arcsin\varepsilon}{\sqrt{1-\varepsilon^2}} 
- \varepsilon\,(1-2\varepsilon^2)\,(3-8\varepsilon^2-8\varepsilon^4)\right]
\\
\qquad\qquad\qquad\times
\left[ \frac{3\,(1-4\varepsilon^2+8\varepsilon^4)\arccos\varepsilon}{\varepsilon} \right.
\\
\left.+ (1-4\varepsilon^2)\,(3+2\varepsilon^2)\sqrt{1-\varepsilon^2}\right].
\label{H-DEDI}
\end{multline} 
where $\varepsilon$ is the dimensionless binding energy,
\begin{equation}
\varepsilon = \frac{\calE}{GM/b}.
\end{equation}

\subsection{Models with a constant anisotropy}
\label{H-conbeta}

The isotropic case can be considered as a special case of the more general class of models with distribution function of the form $f(\calE,L) = f_{\text{A}}(\calE)\,L^{-2\beta}$. These models are characterised by an anisotropy parameter $\beta(r)=\beta$ that is independent of radius. The isotropic case corresponds to $\beta=0$, models with a radial anisotropy have $0<\beta<1$, and models with a preference for tangential orbits have $\beta<0$. Inserting this distribution function in Eq.~(\ref{NE}) and solving the inner integral, one obtains \citep{1991MNRAS.253..414C, 2006PhRvD..73b3524E}
\begin{equation}
N(\calE)
=
\frac{(2\pi)^{5/2}}{2^{\beta-1}}\,\frac{\Gamma(1-\beta)}{\Gamma\left(\frac32-\beta\right)}\,f_A(\calE)
\int_0^{r_\txm(\calE)} \frac{r^{2-2\beta}\,\txd r}{ (\Psi-\calE)^{\beta-1/2}} .
\label{DEDA}
\end{equation}
\citet{2002A&A...393..485B} presented an explicit expression for the distribution function of constant anisotropy Hernquist models in terms of hypergeometric functions, and showed that this distribution function is positive, and hence physical, as long as $\beta\leqslant\tfrac12$. Inserting this expression into equation~(\ref{DEDA}) and evaluating the resulting integral, we find a general expression for the differential energy distribution of the Hernquist model with constant anisotropy $\beta$,
\begin{multline}
N(\calE) 
=
\frac{2b}{G}\,
\frac{\Gamma(3-2\beta)\,\Gamma(5-2\beta)}{\Gamma\bigl(\tfrac92-3\beta\bigr)\,\Gamma\bigl(\tfrac72-\beta\bigr)}\,
(1-\varepsilon)\\\
\qquad\qquad\qquad\times
{}_2F_1\left(\tfrac12-\beta, \tfrac32-\beta, \tfrac92-3\beta, 1-\varepsilon\right)
\\
\times
{}_2F_1\left(-\tfrac32+\beta, \tfrac52+\beta, \tfrac72-\beta, \varepsilon \right).
\label{H-DEDA}
\end{multline}
For all integer and half-integer values of $\beta$, this daunting expression can be written in terms of elementary functions. Setting $\beta=0$ recovers the expression~(\ref{H-DEDI}). A particularly simple case is $\beta=\tfrac12$, the model with the most radial anisotropy that still corresponds to the positive distribution function. The result is \citep[see also][Problem 4.7]{2008gady.book.....B}
\begin{equation}
N(\calE) = \frac{3b}{G}\,(1-\varepsilon)^2.
\end{equation}
The expressions for other integer and half-integer values of $\beta$ are more complex. For example, for $\beta=-\tfrac12$ we have
\begin{multline}
N(\calE) = \frac{b}{3G}\,\frac{(10-10\varepsilon+3\varepsilon^2)}{(1-\varepsilon)^4}
\\
\times
(1-6\varepsilon+18\varepsilon^2-10\varepsilon^3-3\varepsilon^4+12\varepsilon^3\log\varepsilon).
\end{multline}

\subsection{Radial orbit model}

In the limit $\beta\to1$ we have a dynamical model with only radial orbits. As a radial orbit has zero angular momentum, such models are characterised by a distribution function of the form $f(\calE,L) = f_{\text{R}}(\calE)\,\delta(L^2)$.
Inserting this expression into equation~(\ref{NE}), we immediately find
\begin{equation}
N(\calE) 
= 
4\!\sqrt{2}\,\pi^2\,f_{\text{R}}(\calE)
\int_0^{r_\txm(\calE)} \frac{\txd r}{\sqrt{\Psi-\calE}}.
\label{DEDR}
\end{equation}
For the case of the Hernquist model, this case is a bit cumbersome, as the Hernquist model cannot be supported by a model consisting of only radial orbits. Indeed, it has only a $r^{-1}$ cusp, whereas a $r^{-2}$ cusp is required for a radial orbit model to be consistent \citep{1984ApJ...286...27R}. However, we can still formally derive the differential energy distribution for this hypothetical orbital configuration by evaluating the integral~(\ref{DEDR}). The result is
\begin{equation}
N(\calE) = \frac{16b}{15\pi G}\,(5-6\varepsilon)\left[\arccos\!\sqrt{\varepsilon}+\sqrt{\varepsilon\,(1-\varepsilon)}\right].
\label{dedrad}
\end{equation}
This expression can also be obtained by setting $\beta=1$ in the general expression~(\ref{H-DEDA}). This differential energy distribution is negative for $\tfrac56<\varepsilon<1$.

\subsection{Circular orbit model}

Finally, in the limit $\beta\to-\infty$, all stars are on purely circular orbits. The distribution function of such models can be written as \citep{1984ApJ...286...27R}
\begin{equation}
f(\bfx,\bfv) = \frac{1}{\pi}\,\rho(r)\,\delta(v_r)\,\delta(v_t + r\,\Psi').
\end{equation}
Inserting this expression in Eq.~(\ref{NEgen}), we find after some calculation
\begin{equation}
N(\calE) = 4\pi\int_0^\infty \rho(r)\,\delta\Bigl(\Psi+\tfrac12\,r\,\Psi'-\calE\Bigr)\,r^2\,\txd r.
\label{DEDC}
\end{equation}
If the density and the potential are known, this integral can be evaluated and written as a function of $\calE$. For the Hernquist model, this yields the following simple expression
\begin{equation}
N(\calE) = \frac{2b}{G} \left(\frac{3}{\sqrt{1+8\varepsilon}}-1\right).
\end{equation}

\end{document}